\documentclass[12pt]{article}
\usepackage{latexsym} \usepackage{epsf}
\usepackage{a4}
\usepackage{amsfonts}
\usepackage{showlabels}
\usepackage[colorlinks]{hyperref}
\usepackage{color}

\makeatother

\def\be{\begin{equation}}
\def\ee{\end{equation}}
\def\bea{\begin{eqnarray}}
\def\eea{\end{eqnarray}}

\def\'{\prime}

\def\g{\gamma}

\def\ba{\begin{array}}
\def\ea{\end{array}}

\def\nn{\nonumber}
\begin{document}
 {\begin{center} {\LARGE{
Solutions to the Yang-Baxter equations with $osp_q(1|2)$ symmetry: Lax operators } } \\
[8mm] {\large  D. Karakhanyan\footnote{e-mail:
karakhan@mail.yerphi.am} \&
Sh. Khachatryan\footnote{e-mail:shah@mail.yerphi.am} \\
[3mm] } Yerevan Physics Institute , \\
Br.Alikhanian st.2, 375036, Yerevan , Armenia.\\
\vspace{3cm} \noindent {\large{\bf Abstract.}}
\end{center}}

We find a new $4\times4$ solution to the $osp_q(1|2)$-invariant
Yang-Baxter equation with simple dependence on the spectral
parameter and propose $2\times 2$ matrix expressions for the
corresponding Lax operator. The general inhomogeneous universal
spectral-parameter dependent $R$-matrix is derived. It is proven,
that there are two independent solutions to the homogeneous
$osp_q(1|2)$-invariant YBE, defined on the fundamental three
dimensional representations. One of them is the particular case of
the universal matrix, while the second one does not admit
generalization to the higher dimensional cases. Also the $3 \times
3$ matrix expression of the Lax operator is found, which have a
well defined limit at $q \to 1$.

\vspace{5cm}


\newpage
{\tableofcontents}

\section{Introduction}

\setcounter{equation}{0} The simplest orthosymplectic superalgebra
$osp(1|2)$ plays the same role in the classification of Lie
superalgebras \cite{kac} as the algebra $s\ell(2)$ plays in the
classification of Lie algebras. Integrable models with appropriate
non deformed symmetry, based on the solution to the Yang-Baxter
equation (YBE) \cite{ybe}, as well as the quantum deformation of
the superalgebra, $osp_q(1|2)$, were considered in the articles
\cite{k1,k2,k3,bsh,k3,kr,s}. In \cite{k2} it is presented a
fundamental $osp_q(1|2)$-invariant $R(u)$-matrix, found from the
classical matrix \cite{k1} by replacing the rational factors by
trigonometric ones. H.Saleur in 1990 published solution to the
spectral parameter dependent YBE for the universal homogeneous $R$
matrix with the quantum deformation of the $osp(1|2)$ symmetry
\cite{s}. Since that time different authors turned repeatedly
 to the study of the $osp_q(1|2)$ superalgebra and
integrable models with $osp_q(1|2)$ symmetry (see for example
\cite{k,z,ya,m,sca,kp_m,gm,mp,sw,gz}), but some questions are
unclear so far. In particular how many different fundamental $R$
matrices with $osp_q(1|2)$ exist? How to build the Lax operator of
simple form etc.

  In this paper we are constructing the simple form of
the Lax operator and are analyzing the all $osp_q(1|2)$ invariant
solutions to the spectral parameter dependent YBE. For
completeness sake we want to mention the work \cite{g}, where the
authors connect the quest of the Lax operator of $osp_q(1|2n)$
with the isomorphism existing between the representations of
$osp_q(1|2n)$ and $so_q(2n+1)$ \cite{z}. Our investigation in
principle differs from it, as we consider and manipulate directly
the $osp_q(1|2)$ symmetric matrices. Though, of course, the
reminiscence with the $sl_q(2)$ symmetric case can be achieved
along the discussion, and  we shall turn to this equivalence in
the paper (subsection \ref{su2}).

In the next part of the Introduction we give a brief review of the
$osp_q(1|2)$ algebra and it's finite dimensional irreducible
representations (irreps) \cite{k3,kr,s}. In order to determine the
Lax operator, we shall consider YBE defined also for the even
dimensional irreps, which have no classical counterparts
\cite{k,kkh}. The irrep with minimal dimension (higher than one)
is the two dimensional irrep, and in the second section it is
given a simple Lax operator as a $2\times 2$ matrix, defined on
the two dimensional auxiliary space. Using the YB equations with
Lax operators, it is possible to derive determining equations
("Jimbo's relations") for the universal $R$-matrix. It is done in
the third section. The results regard general inhomogeneous
spectral parameter dependent $R$-matrices defined for both the
even and odd dimensional irreducible representations. Note, that
conventional Lax operator, with three dimensional auxiliary space
can be constructed by a descendant procedure (by fusion), from the
$2\times 2$ matrix operator. We calculated the corresponding
$3\times 3$ matrix form of the quantum Lax operator (see fifth
section), and demonstrated the existence of the appropriate
classical limit ($q\to 1$).

Another question, which is analyzed in this work, concerns to the
existence of two fundamental solutions to the YBE (fourth
section). One of them
 follows
from the universal $R(u)$-matrix \cite{s}  as a particular case.
The second solution, existing in the literature \cite{k1,k2,bsh},
seems to be an absolutely separated solution, which does not give
rise to descendant solutions and generalization to a universal
$R(u)$ matrix. Here we examined all the possible solutions to the
spectral parameter dependent YBE,
$$R_{ij}(u-v)R_{ik}(u)R_{jk}(v)=R_{jk}(v)R_{ik}(u)R_{ij}(u-v),$$
with $osp_q(1|2)$ symmetry. We show, that only for the ordinary
fundamental (three-dimensional) representations there are two
different solutions for YBE, while when the dimension of one of
the (irreducible) representations  differs from three, there is
one unique solution $R_{ij}(u)$.

\paragraph{Quantum $osp_q(1|2)$ superalgebra and its
representations.}

 The quantum $osp_q(1|2)$ superalgebra is
defined by the simple odd elements $e,\;f$ and the even element
$h$, obeying the relations
\bea\label{qalg} \{e,f\}=\frac{q^h-q^{-h}}{q-q^{-1}},\;\;
[h,e]=e,\;\;[h,f]=-f,\eea
which at the value $q=1$ degenerate to the relations of
non-deformed superalgebra $osp(1|2)$. The even element $h$
constitutes the Cartan subalgebra.

The definitions of the co-product of the quantum super-algebra are
given by the following relations
\bea \Delta(e)&=&e\otimes q^{-h/2}+q^{h/2}\otimes e,\nn\\
 \Delta(f)&=&f\otimes q^{-h/2}+q^{h/2}\otimes f,\label{cop}\\
 \Delta(h)&=&h\otimes 1+1\otimes h,\nn
 \eea
where $\otimes$ denotes the graded co-multiplication.

 The Casimir operator can be expressed by $e,\; f,\; h$ elements as
  \be\label{cas}c=-(q+2+q^{-1})f^2e^2+
(q^{h-1}+q^{1-h})fe+\big[h-\frac{1}{2}\big]_q^2=\left(
(q^{\frac12}+q^{-\frac12})fe-\left[h-\frac12\right]_q\right)^2.
  \ee
As usual  $[x]_q$ or $[x]$ denotes the expression
$\frac{q^x-q^{-x}}{q-q^{-1}}$.

 The conventional finite
dimensional irreducible representations of this quantum
superalgebra, as well as for the non-deformed case, are odd
dimensional spin-$j$ irreps $V_{4j+1}$, with integer or
half-integer $j$, $dim[\mbox{spin-}j]=4j+1$;
\be V_{4j+1}={\rm span}
\{|j,j\rangle,|j,j-1/2\rangle,...,|j,-j\rangle\}, \quad \;
h|j,i\rangle=2i|j,i\rangle. \label{rep}\ee
The tensor product of two irreps splits into the direct sum of
irreps
\bea V_r\otimes V_k=\bigoplus_{p=|r-k|+1}^{r+k-1} V_p.
\label{vv}\eea

The quantum superalgebra $osp_q(1|2)$ together with the
conventional odd-dimensional spin representations possesses
{\it{even-dimensional}} representations as well \cite{k}, which
have no classical (non-deformed) counterparts, being ill-defined
when $q\to1$. The $r=2n$, $n\in \mathbb{Z}_{+}$, dimensional
representations form a sequence, labelled by positive even integer
$r$ or by "spin"-$j_r$:
  \be\label{lambda}
2j_r=\frac{r-2}{2}+\lambda=2 j-\frac{1}{2}+\lambda,\qquad\qquad
q^\lambda=i q^{\frac{1}{2}},\qquad\qquad \lambda=\frac{i\pi}{2\log
q}+\frac{1}{2}.
  \ee
In particular the "spin" of two-dimensional ($r=2$) representation
is equal to $\lambda/2$. By means of the quarter-integer numbers
$j=j_r-\lambda/2+\frac{1}{4}$, the $4j+1$ dimensional
representations can be written in the same form as the
conventional representations: $V_{4j+1}= {\rm span
}\{|j,j\rangle\,|j,j-1/2\rangle,...,|j,-j\rangle\}$,
 $h|j,i\rangle=2(i+\lambda/2-\frac{1}{4})|j,i\rangle$.

The decomposition of the tensor products, which include even
dimensional representations too, remains the same, as in
(\ref{vv}). Note only, that due to the automorphism of the algebra
$e\to -e,\; f\to f,\; q^{\pm h} \to -q^{\pm h}$, the definition of
the spin-j representations admits an ambiguity: adding the number
$2 n (\lambda-1/2)$, with arbitrary integer $n$, to the  weight
$2j$ doesn't change the irrep \cite{k}. It leads to the
multiplication of the eigenvalues of the operator $q^h$ by the
extra sign $(-1)^n$. It can be seen in the tensor product
(\ref{vv}) of the two even dimensional irreps.

 The eigenvalues of the Casimir operator $c$ on the odd
and even dimensional irreps  $V_p$ are (see \cite{k})
\be\label{ck}
c_{p}=(-1)^{p+1}\frac{q^p+2(-1)^p+q^{-p}}{q^2-2+q^{-2}}=\left\{\ba{cc}
\;[\frac p2]^2,\qquad {\rm if}\;\;p\;\;{\rm is\;\;odd},\\\;[\frac
p2+\lambda]^2,\quad {\rm if}\;\;p\;\;{\rm is\;\;even}.\ea\right.
  \ee
In the space of
 $V_r\otimes V_k$ (\ref{vv}) the Casimir operator is decomposed in terms
 of the projection operators $P_p$ ($P_p$ acts as unity operator
 in the space of the irrep $V_p$ and is $0$ elsewhere)
,
\bea c=\sum_{p= |r-k|+1}^{r+k-1} c_p \;P_p\;,\qquad
\label{p}P_p\cdot V_{p'}=\delta_{p p'}V_{p'}\;.\eea

\paragraph{Intertwiner $R$-matrices and YBE.} Together with the
co-product $\Delta$, the  operation $\bar\Delta=P\Delta P$ is
defined for the quantum (super-)algebras, which differs by
permutation of the spaces: $P\cdot V\otimes V'=V'\otimes V$.
$\Delta$ and $\bar\Delta$ are related by an intertwiner matrix $R$
\be\label{rdelta} R\Delta=\bar{\Delta} R, \ee
which satisfies the so called "constant" Yang-Baxter equations
(YBE) \cite{grs}
\be  R_{12} R_{13} R_{23}= R_{23} R_{13} R_{12}.\ee
Here the expressions in the right and left sides of the equation
act on the tensor product of three vector spaces $V_{r_1}\otimes
V_{r_2}\otimes V_{r_3}$. The $ R_{ij}$ correspondingly acts on the
$V_{r_i}\otimes V_{r_j}$. By a convention, used in the theory of
integrable models, the states $V_{r_1}$ and $V_{r_2}$ are regarded
as "auxiliary" states, while the $V_{r_3}$ state is "quantum" one.

The YBE with spectral parameters
\be  R_{12}(u) R_{13}(u+v) R_{23}(v)= R_{23}(v) R_{13}(u+v)
R_{12}(u). \label{YBR1}\ee
play a crucial role in the theories of the integrable models
\cite{ybe,grs}. If the underlying model has the symmetry of a
quantum (super-)algebra, then the $R(u)$ matrices, with spectral
parameter $u$, satisfy to the relations (\ref{rdelta}). And also
the following relations can be considered,
\be R(u)\Delta_{u}=\bar\Delta_{-u} R(u),\label{rdeltau}\ee
where the operation $\Delta_{u}$ can be treated as a spectral
parameter dependent co-product of the corresponding quantum
(super-)algebra. It has not all the properties of the $\Delta$, it
is not co-associative in particular.
 The
structure of the $\Delta_{u}$ can be checked as in \cite{j} (see
the third section).

 In the so-called "check"-formalism, with check $R$-matrix $\check{R}(u)=P R(u)$,
  $P$ being permutation operator, the following relations are true
 \be\label{symm}
\check{R}(u)\Delta=\Delta\check{R}(u).
 \ee
 This just means that $R$-matrix in "check"-formalism is
commutative with the algebra generators. Hence $\check
R(u)\Delta(c)=\Delta(c)\check R(u)$, where $c$ is Casimir
operator, and $R$-matrix is diagonalizable simultaneously with
Casimir operator and in the space of $V_k\otimes V_r$ can be
represented as a linear combination of the projection operators
introduced in the formula (\ref{p}) \cite{kr}:
\be \check R(u)=\sum_p r_p(u)P_p.\label{rpp} \ee
The "check" $\check{R}(u)$-matrix satisfies to the "check" YBE
%
\be  \check{R}_{12}(u) \check{R}_{23}(u+v) \check{R}_{12}(v)=
\check{R}_{12}(v) \check{R}_{23}(u+v)
\check{R}_{12}(u).\label{YBRk}\ee

As  follows  from (\ref{symm}) and (\ref{YBRk}), the definition of
$R(u)$-matrix has an ambiguity and $R(u)$ can be multiplied by an
arbitrary function.

Note, that for the super-algebras, when writing the tensor
products, co-products and YBE equations in the matrix notations,
one must take into account the grading of the states and
generators. As example, the matrix representation of the tensor
product of two operators $a$ and $b$ reads as
\bea (a\otimes b)_{ij}^{kr}=a_i^k b_j^r
(-1)^{p_k(p_j+p_r)}.\label{tensp}\eea
Here $p_i$ is the parity of the state labelled by $i$ and equals
to $0$ for the even states and equals to $1$ for the odd states.

\section{YBE solution of minimal dimension}
\setcounter{equation}{0}
\subsection{The $4\times 4$ matrix solutions of $RRR=RRR$ YBE}

 Consider now a homogeneous $\check{R}(u)$-matrix on
the tensor product of two two-dimensional representations of the
quantum super-algebra $osp_q(1|2)$. This tensor product is
decomposed to the direct sum of the spaces corresponding to the
conventional one and three dimensional irreps, which means that
the Casimir operator expands into projection operators:
 \bea\label{proj}&c=[\frac32]_q^2P_3+[\frac12]_q^2P_1,&\\
 \nonumber
&\!\!P_1=\frac1{q^{\frac12}-q^{-\frac12}}\left(\!\!\ba{cccc}&&&\\
&-q^{-\frac12}&i&\\ &i&q^{\frac12}&\\&&&\ea\!\!\right)\!\!,\quad
P_3=\frac1{q^{\frac12}-q^{-\frac12}}\left(\!\!\ba{cccc}q^{\frac12}
-q^{-\frac12}&&&\\&q^{\frac12}&-i&\\&-i&-q^{-\frac12}&
\\&&&q^{\frac12}-q^{-\frac12}\ea\!\!\right)\!\!.&
 \eea
In accordance to (\ref{rpp}), the same takes place for $\check
R(u)$ too:
 \be\label{dec}
\check R(u)=a(u)P_1+b(u)P_3,
 \ee
where $a(u)$ and $b(u)$ are some indeterminate functions.

Then we substitute the expression (\ref{dec}) into the Yang-Baxter
relations (\ref{YBRk}), using that in matrix notations  $\check
R(u)_{a_1 a_2}^{b_1 b_2}=R_{a_1 a_2}^{b_2
b_1}(-1)^{p_{b_1}p_{b_2}}$. Written by means of the check matrices
the graded YBE preserve the same matrix representation form as in
the non-graded case, i.e,
\bea \sum_{b_1 b_2 e_2}\check{R}_{a_1a_2}^{b_1 b_2}(u-v)
\check{R}_{b_2 a_3}^{e_2 c_3}(u) \check{R}_{b_1 e_2}^{c_1 c_2}(v)=
\sum_{b_2 e_2 b_3}\check{R}_{a_2 a_3}^{b_2 b_3}(v) \check{R}_{a_1
b_2}^{c_1 e_2} (u)\check{R}_{e_2 b_3}^{c_2 c_3}(u-v).
\label{checkYB} \eea
 The equations
(\ref{checkYB}) with matrix (\ref{dec}) lead to a constraint,
fixing the ratio $a(u)/b(u)$:
$$
\frac{q^{\frac12}b-q^{-\frac12}a}{q^{\frac12}a-q^{-\frac12}b}
=q^{ku},
$$
where the arbitrary constant $k$ can be eliminated by rescaling of
$u$. One can finally set (accurate within multiplication by a
function)
 \be\label{ab}
a(u)=\frac{q^{u-\frac12}+q^{\frac 12-u}}{q^{\frac12}+
q^{-\frac12}}=[u]-[u-1],\qquad b(u)=\frac{q^{u+\frac12}
+q^{-u-\frac12}}{q^{\frac12}+q^{-\frac12}}=[u+1]-[u].
 \ee
It is easy to check then that all the equations in (\ref{checkYB})
turn to be identities upon substitution (\ref{ab}) and one comes
to the following matrix elements of the $R$-matrix:
 \be\label{rmat}
R^{11}_{11}=-R^{22}_{22}=\check R^{11}_{11}=\check R^{22}_{22}
=[u+1]-[u],\qquad R^{21}_{12}=\check R^{12}_{12}=q^{u},
 \ee
$$
R^{12}_{12}=R^{21}_{21}=\check R^{12}_{21}=\check R^{21}_{12}
=-i(q^{\frac12}-q^{-\frac12})[u],\qquad R^{12}_{21}=\check
R^{21}_{21}=q^{-u}.
$$
Finally we can state, that the $R(u)$-matrix (\ref{rmat})
satisfies to graded YBE:
$$
(-1)^{p_{b_2}(p_{a_3}+p_{b_3})}R_{a_1a_2}
^{b_1b_2}(u)R_{b_1a_3}^{c_1b_3}(u+v)R_{b_2b_3}^{c_2c_3}(v)=
(-1)^{p_{b_2}(p_{b_3}+p_{c_3})}R_{a_2a_3}
^{b_2b_3}(v)R_{a_1b_3}^{b_1c_3}(u+v)R_{b_1b_2}^{c_1c_2}(u).
$$
The states with indices $1$ and $2$ have $0$ and $1$ parities
correspondingly: $p_{i}=i-1$.

\subsection{$ RLL=LLR $ YBE: Lax operator}

Here we want to consider the YBE with Lax operator $L(u)$:
 \bea
R^{(22)}(u-v) L^{(2\;r)}(u) L^{(2\;r)}(v)=L^{(2\;r)}(v)
L^{(2\;r)}(u)R^{(22)}(u-v).\label{ybe}\eea
The upper indexes show the dimensions of the auxiliary (dim $=2$)
and quantum (dim $=r$, arbitrary integer) states on which
$R$-matrix and $L$-operator act. The matrix elements of
$R^{(22)}(u)$ are defined in (\ref{rmat}).

The experience with solutions to Yang-Baxter equations with
$gl(n)$-symmetry (or more generally with symmetry with respect to
the superalgebra $gl(m|n)$) or their quantum deformations suggests
us that solution corresponding to the representation with the
smallest dimension in auxiliary space has the simplest (linear)
dependence on spectral parameter $u$ ($[u]_q$ in quantum case).
The power of $u$ grows alone with the growth of dimension of the
representation in the auxiliary space. Supposing that this
regularity can be continued to the more complicated algebras, one
can try the following $2\times 2$ matrix as a solution to the
graded YBE (\ref{ybe}):
 \be\label{lax}
L(u)=\left(\ba{cc}[u-\lambda h+1]-[u-\lambda h]&(q^{\frac12}-
q^{-\frac12})^{\frac12}q^{-u}f\\-i(q^{\frac12}-q^{-\frac12})^{
\frac12}q^{u}e&[u-(\lambda-1)h]-[u-(\lambda-1)h+1]\ea\right),
 \ee
where $e$, $f$ and $h$ are the elements of $osp_q(1|2)$ acting
 on
the $r$-dimensional space. The complex number $\lambda$ is defined
by  the relation $[\lambda-1]= [\lambda]$, or $
q^\lambda=iq^{\frac12}$.

 In terms of matrix elements (\ref{ybe}) takes the form:
 \be\label{rll}
(-1)^{p_{b_2}(p_{b_1}+p_{c_1})}R_{a_1a_2}
^{b_1b_2}(u-v)L_{b_1}^{c_1}(u)L_{b_2}^{c_2}(v)=
(-1)^{p_{b_2}(p_{a_1}+p_{b_1})}
L_{a_2}^{b_2}(v)L_{a_1}^{b_1}(u)R_{b_1b_2}^{c_1c_2}(u-v).
 \ee
The inspection shows that the equations (\ref{rll}) are satisfied
based on the algebra relations (\ref{qalg}) and on two identities:
 \be\label{i1}
([u-v+1]-[u-v])([u+x+1]-[u+x])=[v+x+1]-[v+x]+(q^{\frac12}-
 q^{-\frac12})^2[u-v][u+x+1],
 \ee
and
 \be\label{i2}
([u-v+1]-[u-v])([v+x+1]-[v+x])=[u+x+1]-[u+x]-(q^{\frac12}-
 q^{-\frac12})^2[u-v][v+x].
 \ee
\section{ General Lax operator and Jimbo's relations; connection with the $sl_q(2)$ }
\setcounter{equation}{0}
\subsection{General Lax operator and Jimbo's relations; $\check{R}^{(r_1 r_2)}(u)$}

In this subsection we want to observe the YBE
\bea R^{(r_1 r_2)}(u-v) L^{(2\;r_2)}(u)
L^{(2\;r_1)}(v)=L^{(2\;r_1)}(v) L^{(2\;r_2)}(u)R^{(r_1
r_2)}(u-v),\label{ybe1}\eea
with $L^{(2\;r)}(u)$ and a general $R^{(r_1 r_2)}(u)$, defined on
the $r_1$ and $r_2$ dimensional spaces.

In terms of the check $\check{R}$-matrix elements the equation
(\ref{ybe1}) takes the form
\bea\nn & (-1)^{p_{c_1}(p_{b_2}+p_{c_2})}\check{R}_{a_1a_2}
^{b_1b_2}(u-v)(L_{\alpha}^{\beta})_{b_1}^{c_1}(u)(L_{\beta}^{\gamma})_{b_2}^{c_2}(v)=
\qquad\qquad\qquad\qquad\qquad&\\
&\qquad\qquad\qquad\qquad\qquad= (-1)^{p_{b_1}(p_{a_2}+p_{b_2})}
(L_{\alpha}^{\beta})_{a_1}^{b_1}(v)
(L_{\beta}^{\gamma})_{a_2}^{b_2}(u)\check{R}_{b_1b_2}^{c_1c_2}(u-v).&
\label{ybec1}\eea
Here it is used the condition
$(-1)^{p_{a_1}+p_{a_2}+p_{b_1}+p_{b_2}}=1$ for the $R^{a_1a_2}
_{b_1b_2}$. The Greek indexes relate to the $2$-dimensional
quantum space, while the Latin ones to the $r_1,\; r_2$
dimensional auxiliary spaces. Recalling the definition
(\ref{tensp}), we can trace the graded tensor products in the
l.h.s and r.h.s of the equation (\ref{ybec1}).
\paragraph{General form of the Lax operator.}
 At firs let us present the Lax operator $L^{(2\;r)}$ in rather general
form
\be \label{g} L(u)=\left(\ba{cc} a_1(u) q^{\lambda h}+ a_2(u)
q^{-\lambda h}&g_1(u)f
\\g_2(u)e&c_1(u) q^{(\lambda-1) h}+ c_2(u) q^{-(\lambda-1)
h}\ea\right).
 \ee
It follows from the YBE (\ref{ybe}) that the coefficient functions
and $\lambda$ must satisfy the
 relations
\bea \nn &q^\lambda=i q^{1/2},\qquad g_1(u)=a_1(u) g_{10},
&g_2(u)=a_1(u)q^{-2 u}g_{20},\\
&a_2(u)= a_1(u) q^{-2 u} a_{20},\quad c_1(u)=a_1(u)c_{10},
&c_2(u)=a_1(u)q^{-2u}a_{20}c_{10},\label{acg} \eea
with one constraint,
\be\frac{g_{10}g_{20}}{a_{20}c_{10}}=-i\frac{(1+q)^2(q-1)}{q^{3/2}},\label{gacq}\ee
on the constant coefficients $ g_{20},g_{10},a_{20},c_{10}$.

The particular solution given by the matrix (\ref{lax})
corresponds to a symmetric choice of the constants
$g_{20},g_{10}$,  with preliminarily change  $q\to q^{-1}$,
$$a_1(u)=\frac{q^{u+1}}{1+q},\;a_{20}=1/q,\;\;c_{10}=-1,\;\;
g_{20}=-i g_{10}=i\frac{(q-1)^{1/2}(1+q)}{q^{5/4}}.$$

So, taking into account all the constraints (\ref{acg},
\ref{gacq}), the formula (\ref{g}) takes the form
 \be\label{lgg}
L(u)=a_1(u) q^{-u}\left(\ba{cc} q^{u+\lambda h}+a_{20}q^{-
u-\lambda h}
 &q^{u} g_{10}f
\\q^{-u} g_{20}e&c_{10} (q^{u+(\lambda-1) h}+a_{20}q^{u-(\lambda-1)
h})\ea\right).
 \ee
Before proceeding further, let us to analyze the cause of the
existence of the arbitrary constants $g_{10},a_{20},c_{10}$. As it
was stated before, YBE allow an arbitrariness in solutions in the
form of the multiplier functions, hence the appearance of the
function $a_1(u)$. Using the constraint (\ref{gacq}), and after
some manipulations $L(u)$ takes the form
\be L(u)=a_1(u)(a_{20})^{\frac{1}{2}} q^{-u}\left(\ba{cc}
q^{u'+\lambda h}+q^{- u'-\lambda h}
 &q^{u'} g_{10}f
\\-i
\frac{(1+q)^2(q-1)}{q^{3/2}}q^{-u'}\frac{c_{10}}{g_{10}}e\;\;\;
&c_{10} (q^{u'+(\lambda-1)h} +q^{u'-(\lambda-1) h})\ea\right),
 \ee
with $q^{u'}=q^u/(a_{20})^{\frac{1}{2}}$. So, the constant
$a_{20}$ can be eliminated by redefinition of the spectral
parameter. It is obvious also, that arbitrary constant $g_{10}$
has appeared due to the known automorphism $f\to g_{10}f, \; \; \;
g\to e/g_{10}$ of the algebra. It remains only one constant,
$c_{10}$, the nature of which will be apparent in the next part of
this section (see (\ref{copc1})).
\paragraph{Jimbo's relations.}
Choosing $a_1(u)=q^{u}$, let us rewrite (\ref{lgg}) as
 \bea\nn
&L(u)=q^u L_{+}+q^{-u}L_{-}\;,&
\\&L_{+}=\left(\ba{cc} q^{\lambda
h}&g_{10}f
\\0&c_{10}q^{(\lambda-1) h}\ea\right),&
\\\nn
&L_{-}=\left(\ba{cc} a_{20} q^{-\lambda h}&0
\\g_{20}e&c_{10} a_{20} q^{-(\lambda-1)
h}\ea\right).&
 \eea
 This separation transforms the YBE (\ref{ybec1}) to a few relations,
 below we represent the crucial ones:
 \bea\label{jimbo1}
 \check{R}(u)(L_{\pm})_\alpha^\beta{\otimes}(L_{\pm})_\beta^\gamma=
 (L_{\pm})_\alpha^\beta{\otimes}(L_{\pm})_\beta^\gamma\check{R}(u).
 \eea
\be\label{jimbo2} \check{R}(u)\{q^u (L_{+})_\alpha^\beta\otimes
(L_{-})_\beta^\gamma+q^{-u} (L_{-})_\alpha^\beta\otimes
(L_{+})_\beta^\gamma\}=\{q^{-u} (L_{+})_\alpha^\beta\otimes
(L_{-})_\beta^\gamma+q^u (L_{-})_\alpha^\beta\otimes
(L_{+})_\beta^\gamma\}\check{R}(u).\ee
In (\ref{jimbo1}, \ref{jimbo2}) the summation of the products by
the Latin indexes is replaced by the graded tensor product
operation symbol. From the equations (\ref{jimbo1}) the symmetry
relations (\ref{cop}, \ref{symm}) follow, as can be verified.
 The
 equations
\bea \label{rhh}&\check{R}(u)(q^{\pm \lambda h}\otimes q^{\pm
\lambda h})=(q^{\pm \lambda h}\otimes q^{\pm \lambda
h})\check{R}(u),&\\\nn &\check{R}(u)(q^{\pm (\lambda-1) h}\otimes
q^{\pm (\lambda-1) h})=(q^{\pm(\lambda-1) h}\otimes
q^{\pm(\lambda-1) h})\check{R}(u),&\eea
appear in the diagonal parts of (\ref{jimbo1}). As their
consequences the following relations are derived
\be\check{R}(u)(q^{\pm h}\otimes q^{\pm h})=(q^{\pm h}\otimes
q^{\pm h})\check{R}(u).\ee
And similarly, from the next equations,
\bea &\check{R}(u)(q^{\lambda h}\otimes f+c_{10}f\otimes
q^{(\lambda-1)h})=(q^{\lambda h}\otimes f+c_{10}f\otimes
q^{(\lambda-1)h})\check{R}(u),&\\\nn &\check{R}(u)(e\otimes
q^{-\lambda h}+c_{10}q^{-(\lambda-1)h}\otimes e)=(e\otimes q^{
-\lambda h}+c_{10}q^{-(\lambda-1)h}\otimes e)\check{R}(u),&\eea
constituting the non-diagonal parts of the (\ref{jimbo1}), the one
can verify, using (\ref{rhh}), that the following relations are
derived,
\bea \label{copc}&\check{R}(u)(q^{h/2}\otimes f'+c_{10}f'\otimes
q^{-h/2})=(q^{h/2}\otimes f'+c_{10}f'\otimes
q^{-h/2})\check{R}(u),&\\\nn &\check{R}(u)(e'\otimes q^{
-h/2}+c_{10}q^{h/2}\otimes e')=(e'\otimes q^{
-h/2}+c_{10}q^{h/2}\otimes e')\check{R}(u),& \eea
with corresponding redefinitions $f'=q^{-(\lambda-1/2)h}f,\;\;e'=e
q^{(\lambda-1/2)h}$, which are possible due to the automorphisms
of the algebra. The number $c_{10}$ is arbitrary. When it equals
to $1$, the relations (\ref{copc}) are equivalent to (\ref{cop}).
Though the new definitions for the co-products
\bea\Delta^c(f')\!=\!\frac{1}{\sqrt{c_{10}}}\left(q^{h/2}\otimes
f'+c_{10}f'\otimes q^{-h/2}\right), \quad\!\!
\Delta^c(e')\!=\!\frac{1}{\sqrt{c_{10}}}\left(e'\otimes q^{
-h/2}+c_{10}q^{h/2}\otimes e'\right)\!,\label{copc1}\eea
 where $c_{10}$ is arbitrary, are  allowed.

From equations (\ref{jimbo2})  we obtain the following relations
\bea \nn& \check{R}(u)(c_{10} q^u f\otimes
q^{-(\lambda-1)h}+q^{-u} q^{-\lambda h}\otimes f)=(c_{10}q^{-u}
f\otimes q^{-(\lambda-1)h}+q^u q^{-\lambda h}\otimes
f)\check{R}(u),&\\\nn &\check{R}(u)(c_{10}q^u
q^{(\lambda-1)h}\otimes e+q^{-u} e\otimes q^{\lambda
h})=(c_{10}q^{-u} q^{(\lambda-1)h}\otimes e+q^u f\otimes
q^{\lambda h})\check{R}(u).& \eea
After some easy calculations, taking into account, that
$q^{(2\lambda-1)h}f'=(-1)^{h}f'$ and
$e'q^{-(2\lambda+1)h}=e'(-1)^{h}$, we come to the following
relations (called "Jimbo's relations", \cite{j}),
\bea\nn&\check{R}(u)\left(c_{10} q^u (-1)^{h}f'\otimes
q^{h/2}+q^{-u} q^{-h/2}\otimes
(-1)^{h}f'\right)=\quad\quad\quad\quad\quad\quad\quad\quad\quad\quad&\\\nn&
\quad\quad\quad\quad\quad\quad\quad\quad\quad\quad=\left(c_{10}q^{-u}(-1)^{h}f'\otimes
q^{h/2}+q^u q^{-h/2}\otimes
(-1)^{h}f'\right)\check{R}(u),&\\\label{jimbo3}
&\check{R}(u)\left(c_{10}q^u q^{-h/2}\otimes e'(-1)^{h}+q^{-u}
e'(-1)^{h}\otimes
q^{h/2}\right)=\quad\quad\quad\quad\quad\quad\quad\quad\quad\quad&\\\nn
&\quad\quad\quad\quad\quad\quad\quad\quad\quad\quad
=\left(c_{10}q^{-u} q^{-h/2}\otimes e'(-1)^{h}+e'(-1)^{h}\otimes
q^{h/2}\right)\check{R}(u).& \eea
The operators in the brackets can be treated as definitions of the
co-products, dependent on the continuous parameter $u$:
\bea \Delta^c_u(f')=\frac{1}{\sqrt{c_{10}}}\left(c_{10} q^u
(-1)^{h}f'\otimes q^{h/2}+q^{-u} q^{-h/2}\otimes
(-1)^{h}f'\right),\\
\nn\Delta^c_u(e')=\frac{1}{\sqrt{c_{10}}}\left(c_{10}q^u
q^{-h/2}\otimes e'(-1)^{h}+q^{-u} e'(-1)^{h}\otimes
q^{h/2}\right). \eea
With these definitions the formula (\ref{rdeltau}) is valid, as it
follows from (\ref{jimbo3}).

In the homogeneous case, when two irreps in the tensor product
have the same dimension, the generators can be redefined
such that
$e'(-1)^{h},\;\;(-1)^{h}f'$ operators are replaced by
$e'(-1)^{p},\;\;(-1)^{p}f'$, where $p$ is to define the parity of the
state.

 One can verify, that the
following equations hold
\bea &\left\{(-1)^{h}f'\otimes
q^{h/2},\Delta^c(f')\right\}=0,\qquad &\left\{q^{-h/2}\otimes
(-1)^{h}f',\Delta^c(f')\right\}=0,\\
&\left\{q^{-h/2}\otimes e'(-1)^{h},\Delta^c(e')\right\}=0,\qquad
&\left\{e'(-1)^{h}\otimes q^{h/2},\Delta^c(e')\right\}=0.
 \eea
The consequences of the above relations are particularly the
observations:
\bea \!e'(-1)^{h}\otimes
q^{h/2}|j,j\rangle=\gamma(j)|j+\!\frac{1}{2},j+\!\frac{1}{2}\rangle,\quad
q^{-h/2}\otimes
e'(-1)^{h}|j,j\rangle=\bar{\gamma}(j)|j+\!\frac{1}{2},j+\!\frac{1}{2}\rangle.
\eea

 Let us follow  the procedure proposed in the
 paper \cite{j} and  find the matrix ${\check{R}}^{(r_1 r_2)}(u)$,
 which acts on the product of the spaces $V_{4i_1+1}\otimes
V_{4i_2+1}$, $r_k=4 i_k+1$. The homogeneous case restricted to the
odd dimensional conventional irreps was calculated in the work
\cite{s}. First we write ${\check{R}}^{(r_1 r_2)}(u)$ in the
general form
\be {\check{R}}^{(r_1
r_2)}(u)=\sum_{j=|i_1-i_2|}^{i_1+i_2}\mathbf{r}_j(u)\breve{P}_{4j+1},\label{rch}\ee
with projector operators $\breve{P}_r,\;\breve{P}_r \cdot
V_{g}=\delta_{r g}V_g$, acting as map $V_{4 i_1+1}\otimes V_{4
i_2+1}\to V_{4i_2+1}\otimes V_{4i_1+1}$. When $r_1=r_2$, then
$\check{P}_r=P_r$. And then we place $\check{R}(u)$ in
(\ref{jimbo3}). The action of the r.h.s and the l.h.s of the
equation (\ref{jimbo3}) on the state $|j,j\rangle$ gives the
wanted recurrent equation for the $\mathbf{r}_j(u)$ functions:
\bea \mathbf{r}_{j+\frac{1}{2}}(u)\left(c_{10} q^u
\bar{\gamma}_{i_1 i_2}(j)+q^{-u} \gamma_{i_1
i_2}(j)\right)=\left(c_{10}q^{-u}\bar{\gamma}_{i_2 i_1}(j)+q^u
\gamma_{i_2 i_1}(j)\right)\mathbf{r}_j(u).\eea
The appearance of the indexes $i_1 i_2$ and $i_2 i_1$ in the
relation shows, that there is a difference between the action of
an operator product $a\otimes b$ on the space $V_{r_1}\otimes
V_{r_2}$ and the action on the space $V_{r_2}\otimes V_{r_1}$,
when $r_1\neq r_2$.

For finding a connection between the $\gamma$ coefficients we are
using the relations
\bea \nn&((-1)^{h}\otimes q^{h})\Delta^c(e')|j,j\rangle=&\\
&\frac{1}{\sqrt{c_{10}}}\left((-1)^{h}e'\otimes
q^{h/2}+c_{10}((-1)^{h}q^{h}\otimes (-1)^{h}q^{h}) q^{-h/2}\otimes
(-1)^{-h}e'\right))|j,j\rangle=&\\\nn
&-\frac{1}{\sqrt{c_{10}}}\left(\gamma_{i_1
i_2}(j)+(-1)^{\mathbf{p}_0^{i_1 i_2}}c_{10}(-q)^{h_j+1}
\bar{\gamma}_{i_1 i_2}(j)\right) |j,j\rangle=0.&\eea
Here the values of $h_j$ are the eigenvalues of the $h$ operator
on the highest vector of an irrep; when dimension $r$ is odd, they
coincide with $2j$, $h_j\equiv\frac{r-1}{4}$ (\ref{rep}), and have
an additional complex summand in the even case, as we have seen in
(\ref{lambda}), i.e $h_j=2j_r\equiv 2j+\lambda-1/2$.
$\mathbf{p}_0^{i_1 i_2}=0$ for the case, when both of $r_1$ and
$r_2$ are either even or odd numbers, i.e., when $|r_1-r_2|$ is an
even number. Otherwise, when the second irrep in the tensor
product is even dimensional, $\mathbf{p}_0^{i_1 i_2}=-2\lambda$.
So, the recurrent relations are solved as
\bea\nn
 \mathbf{r}_{j'}(u)=\prod_{j=j'}^{i_1+i_2-1/2}
 \frac{\left(c_{10} q^u \bar{\gamma}_{i_1
i_2}(j)+q^{-u} \gamma_{i_1
i_2}(j)\right)}{\left(c_{10}q^{-u}\bar{\gamma}_{i_2 i_1}(j)+q^u
\gamma_{i_2 i_1}(j)\right)}\mathbf{r}_{i_1+i_2}(u)\\\label{rij}
=\prod_{j=j'}^{i_1+i_2-1/2}\frac{\gamma_{i_1 i_2}(j)}{\gamma_{i_2
i_1}(j)}\frac{\left(q^{u}- q^{-u}(-1)^{\mathbf{p}_0^{i_1
i_2}}(-q)^{2 h_j+1}\right)}{\left(q^{-u}-q^{u}(-1)^{\mathbf{p}_0^{
i_2 i_1}}(-q)^{2h_j+1}\right)}\mathbf{r}_{i_1+i_2-1}(u).
 \eea

$\bullet$ For freeing the formulas in the brackets (\ref{rij})
from the unwanted factors like $(-1)^{x\lambda}$, which arise when
$|r_1-r_2|$ is odd number, we can redefine the spectral parameter
$u$ by an appropriate shift. Suppose $i_1$-irrep has odd
dimension, while the second irrep, with "spin" $i_2$, has even
dimension. Then $\mathbf{p}_0^{i_1
i_2}\equiv\mathbf{p}_0=-2\lambda$, and $\mathbf{p}_0^{i_2 i_1}=0$.
Now all the $h_j$-s represent even dimensional irreps and are
complex. Using the quarter integer numbers $j$ introduced in
(\ref{lambda}) and also recalling that $(-q)=q^{2\lambda}$,
$(-1)=q^{2\lambda-1}$, it is possible to redefine the formulas in
 (\ref{rij}) as follows
\bea
 \mathbf{r}_{j'}(u')=\prod_{j=j'}^{i_1+i_2-1/2}
 \frac{\gamma_{i_1
i_2}(j)}{\gamma_{i_2 i_1}(j)}q^{2\lambda(\lambda-1/2)}
\frac{\left( q^{u'}-(-q)^{2j}q^{-u'}
\right)}{\left(q^{-u'}-(-q)^{2j}q^{u'} \right)}
\mathbf{r}_{i_1+i_2}(u'),\eea
where $u'=u+\lambda(\lambda-1/2)$. $\bullet$

In the homogeneous case, when $i_1=i_2$, and hence
$\frac{\gamma_{i_2 i_1}(j)}{\gamma_{i_1 i_2}(j)}=1$, the
expression (\ref{rij}) coincides with a similar expression,
derived in \cite{s} for odd dimensional irreps. Now, for obtaining
the fraction $\frac{\gamma_{i_2 i_1}(j)}{\gamma_{i_1 i_2}(j)}$ in
general inhomogeneous cases, we shall express the coefficients
$\gamma_{i_1 i_2}(j)$ by means of the Clebsh-Gordan coefficients
\cite{kkh,s}. Suppose (see \cite{kkh})
\bea &|j,j\rangle_{12}=\sum_{j_1 +j_2=j}C\left(^{i_1 i_2 j} _{j_1
j_2 j}\right)|i_1,j_1\rangle \otimes |i_2,j_2\rangle,&\\
&|j,j\rangle_{21}=\sum_{j_1 +j_2=j}C\left(^{i_2 i_1 j} _{j_2 j_1
 j}\right)|i_2,j_2\rangle \otimes |i_1,j_1\rangle,&\\
&e'|i,j\rangle=\alpha_{i}^{2j}|i,j+1/2\rangle.& \eea
Then we have (we assume, that $i_2\geq(j+\frac{1}{2}-i_1)$)
\bea &e'(-1)^{h}\otimes
q^{h/2}|j,j\rangle_{12}=\alpha_{i_1}^{2i_1-1}(-1)^{2
i_1-1}q^{j-i_1+1/2}\frac{C\left(^{\;\; i_1\;\;\;\;\;\;
i_2\;\;\;\;\;\;\;\; j} _{ i_1-\frac{1}{2}\; j+\frac{1}{2}-i_1\;
j}\right)}{C\left(^{i_1\;\;\;\;\;\; i_2\;\;\;\;\;\;j+\frac{1}{2}}
_{i_1\;
j+\frac{1}{2}-i_1\;j+\frac{1}{2}}\right)}|j+\!\frac{1}{2},j+\!\frac{1}{2}\rangle_{12},&\\
&q^{-h/2}\otimes
e'(-1)^{h}|j,j\rangle_{12}=\alpha_{i_2}^{2(j-i_1)}(-1)^{2
(j-i_1)+p_{i_1}}q^{-i_1}\frac{C\left(^{i_1\;\;\; i_2\;\; j} _{i_1
j-i_1\;j}\right)}{C\left(^{i_1\;\;\;\;\; i_2\;\;\;\;\;\;
j+\frac{1}{2}} _{i_1\;
j+\frac{1}{2}-i_1\;j+\frac{1}{2}}\right)}|j+\!\frac{1}{2},j+\!\frac{1}{2}\rangle_{12}.&\\
&\frac{\gamma_{i_2 i_1}(j)}{\gamma_{i_1
i_2}(j)}=(-1)^{2(i_2-i_1)}q^{i_1-i_2}\frac{\alpha_{i_2}^{2i_2-1}}{\alpha_{i_1}^{2i_1-1}
}\frac{C\left(^{i_1\;\;\;\;\;\; i_2\;\;\;\;\;\;j+\frac{1}{2}}
_{i_1\; j+\frac{1}{2}-i_1\;j+\frac{1}{2}}\right)C\left(^{\;\;
i_2\;\;\;\;\;\; i_1\;\;\;\;\;\;\;\;\; j} _{ i_2-\frac{1}{2}\;
j+\frac{1}{2}-i_2\; j}\right)}{C\left(^{\;\; i_1\;\;\;\;\;\;
i_2\;\;\;\;\;\;\;\;\; j} _{ i_1-\frac{1}{2}\; j+\frac{1}{2}-i_1\;
j}\right)C\left(^{i_2\;\;\;\;\;\; i_1\;\;\;\;\;\;j+\frac{1}{2}}
_{i_2\; j+\frac{1}{2}-i_2\;j+\frac{1}{2}}\right)}.&
 \eea
%

\paragraph{Some examples.}
Here we represent the exact non-homogeneous check $R$-matrices for
some simple cases.
\bea
&\check{R}^{(23)}(u)=\check{P}_4+\left(\frac{1}{q^{1/2}-q^{-1/2}}\right)\frac{(q^{-u}+q^{3/2+u})}{(q^u-q^{3/2-u})}
\check{P}_2,&\\
&\check{R}^{(24)}(u)=\check{P}_5+\left(\frac{q^{1/2}+q^{-1/2}}{q^{3/2}+q^{-3/2}}\right)
\frac{(q^{-u}+q^{2+u})}{(q^{u}+q^{2-u})}\check{P}_3,&\\
&\check{R}^{(35)}(u)=\check{P}_7+\left(\frac{1}{q+q^{-1}}\right)
\frac{(q^{-u}-q^{3+u})}{(q^{u}-q^{3-u})}\check{P}_5+\left(\frac{1}{q+q^{-1}}
\frac{q^{1/2}+q^{-1/2}}{q^{3/2}+q^{-3/2}}\right)\frac{(q^{-u}-q^{3+u})(q^{-u}+q^{2+u})}
{(q^{u}-q^{3-u})(q^{u}+q^{2-u})} \check{P}_3.&\eea
\subsection{Connection with  the quantum algebra $sl_q(2)$ \label{su2}}
 \paragraph{Representations and universal $R$-matrices.}
 In \cite{k} the author has demonstrated a correspondence between
 $sl_q(2)$ and  $osp_q(1|2)$ algebras at the level of finite
 dimensional representations and universal $R$-matrices
 (see also \cite{kr, z}). Below we give a
 brief overview of that study.

 If
 $e,\;f,\;h$ are the two-dimensional $osp_q(1|2)$ generators, then
 $E,\;F,\;H$
 \bea
 E=-i(q^{1/2}-q^{-1/2})e,\quad
 F=\left(\ba{cc}1&0\\0&-1\ea\right)f,\quad
 H=2h-2(\lambda-1/2).
 \eea
generate the quantum algebra $sl_t(2)$, with deformation parameter
$t=i q^{1/2}$.

For the general even dimensional representations the
correspondence is given by the relations:
\bea \label{even}
 E=-(t+t^{-1})e,\quad
 F=(-i)i^{2h-\lambda+1/2}f,\quad H=2h-2(\lambda-1/2).
 \eea
In the odd dimensional case the correspondence is stated by the
formulae
\bea\label{odd}
 E=-(t+t^{-1})e,\quad
 F=(-1)^{h}f,\quad H=2h.
 \eea

The universal $R$-matrix for the $sl_t(2)$ algebra is found by
simple replacements in the $osp_q(1|2)$ invariant universal
$R$-matrix
\bea\label{runiv-q} && R_{osp_q(1|2)}=q^{h\otimes
h}\sum_{n=0}^{\infty}\frac{(-q^{1/2})^{1/2n(n-1)}(q-q^{-1})^n}{[n]_{+}!}q^{-nH/2}
e^n\otimes f^n q^{nH/2},\\
&& {[n]_{+}}=\frac{(-1)^{n-1}q^{n/2}+q^{-n/2}}{q^{1/2}+q^{-1/2}},
 \eea
after inserting the expressions in (\ref{even}) or (\ref{odd}),
and replacing $i q^{1/2}$ factors by $t$.

The resulting matrix $R_{sl_t(2)}$
 is understood as matrix
representation. If to take into account the fermionic character of
the tensor products by using appropriate signs for each
representation, the extra factors $(-1)^{H\otimes H}$ and $(-1)^{n
H}$ would be cancelled in the $R_{sl_t(2)}$ and it will be the
universal $R$-matrix of $sl_t(2)$.

It is noted in \cite{k}, that there are slightly different
formulae for stating the correspondence (\ref{even}, \ref{odd})
(see references in \cite{k}). A correspondence of $R$-matrices for
the odd dimensional irreps is discussed in \cite{s}, there the
correspondent matrices are differing by a gauge transformation.
The \cite{z} devotes to the correspondence of the conventional odd
dimensional (non-spinorial) representations of the $osp_q(1|2n)$
and $so_q(2n+1)$.

\paragraph{$R(u)$ matrices and Lax operators.}

Now a correspondence with $sl_t(2)$ case can be stated also for
the Lax operators and $R$-matrices (\ref{rmat}, \ref{lgg},
\ref{rch}), constructed in the first sections.

For example, let us represent a general form of the Lax operator
in terms of $E,\;F,\; H$ generators (\ref{odd}), defined for
conventional irreps.
\be \label{grs} L_{t}(u)=\left(\ba{cc} a_1(u,t) t^{H/2}+ a_2(u,t)
t^{- H/2}&-g_1(u,t)\frac{F}{t+t^{-1}}
\\g_2(u,t)(-1)^{-h}E&c_1(u,t)t^{H/2} (-t)^{-2h}+ c_2(u,t) t^{-H/2}(-t)^{
2h}\ea\right).
 \ee
 There are relations among the functions $a_i(u,t),\;
 c_i(u,t),\;g_i(u,t)$, given in the equations (\ref{acg},
 \ref{gacq}).
 The operator (\ref{grs}) satisfies the graded Yang-Baxter
relations, with the four dimensional $R$-matrix, where one also
has to replace the $q$ parameter by $t$. One can arrive at the
ordinary $sl_t(2)$ Lax operators if one attributes the $(-1)^{h}$
type factors in (\ref{grs}) to the graded character of the
original
 operator and YB equations.

\section{All  solutions to YBE  with  higher dimensional irreps}
\setcounter{equation}{0}

\subsection{Customary
(standard) three dimensional fundamental representations}

 The fundamental representation of the $osp(1|2)$ is the
three dimensional one. There are two known solutions to the
spectral parameter dependent YBE equations (\ref{YBR1}), with
$osp_q(1|2)$ symmetry, which can be found from the articles
\cite{k2} and \cite{s}.

Let us find all the solutions to  (\ref{YBR1}) with $9\times 9$
$R$-matrices,  acting on the product of the fundamental irreps
$V_3\otimes V_3$. Remind, that the $R$-matrices, being the
intertwiner matrices for the $osp_q(1|2)$ superalgebra, satisfy to
(\ref{symm}), which implies the decomposition (\ref{rpp}). So we
are looking for $\check{R}$ in the form
 \be\check{R}^{(33)}(u)=P_5+f(u)P_{3}+g(u)P_{1},\label{rp}\ee
where $P_i$-s are the projectors acting on the decomposition
$V_3\otimes V_3=V_1\oplus V_3\oplus V_5$.

 $R$ is defined within the multiplication by a function of the
spectral parameter. Here and henceforth we choose that function
so, that $\check{R}(0)=1$.

Now let us to turn to the investigation of the equations
(\ref{checkYB}). It turns out, that all the possible solutions can
be found by considering only three independent equations from the
set (\ref{checkYB}). Inserting (\ref{rp}) representation in the
mentioned  equations and taking at first the equation with
elements
$$\{a_1,a_2,a_3, c_1,c_2,c_3\}=\{1,2,1,2,1,1\},$$ we found a
functional relation for the $f(u)$,
\bea f(v)=\frac{(1+q^4)f(u)+q^2(1-f(u-v)+f(u)+f(u-v)f(u))}
{(1+q^4)f(u-v)+q^2(1+f(u-v)-f(u)+f(u-v)f(u))}. \eea
Solving this relation, we found that
\be f(0)=1,\qquad f(u)f(-u)=1,\qquad f(u)=\frac{q^{u
p}q^2-1}{q^2-q^{u p}}.\ee
The constant $p$ is arbitrary, since the transformations $u\to p u,\;
v\to pv$ leave YBE invariant.

As the next equations we take $$\{a_1,a_2,a_3,
c_1,c_2,c_3\}=\{2,2,1,1,2,2\}\;\;\; \mbox{and}\;\;\;
\{a_1,a_2,a_3, c_1,c_2,c_3\}=\{2,2,1,2,2,1\}.$$
The first of them gives an expression for the $g(u)$ function,
$g(0)=1$, without fixing $g'(0)$,
\bea
g(u)=\frac{g'(0)(q\!-\!1)(q^{up}\!-\!1)(q^{4+up}\!-\!1)-p(1\!-\!q+q^2)(1+(q+q^2\!-\!2\!-\!2q^3)q^{up}+q^{2up+3})}
{(q^2\!-\!q^{up})(g'(0)q(q\!-\!1)(q^{up}\!-\!1)-p(1\!-\!q+q^2)(q^{up}\!-\!q))},\eea
 and the second equation gives
two possibilities for the $g'(0)$,
\bea g'(0)=2p\frac{1-q+q^2}{q^2-1} \quad\mbox{(a)}\qquad
{\textrm{and}}\qquad
g'(0)=p\frac{1+q^3}{q^3-1}\quad\mbox{(b)},\label{33}\eea
corresponding to the two known solutions. After substitution the
expressions (\ref{33}) into the formula of $g(u)$ and fixing the
values of $p$, we recover Saleur's solution, (\ref{33}, a),
 \be\label{sal}
{\check R}_{a}(u)=P_5+\frac{1-q^2y}{y-q^2}P_{3}+ \frac{(1+q
y)(1-q^2y)}{(y+q)(y-q^2)}P_1,\qquad y=q^{-2u},
 \ee
and Kulish's solution, (\ref{33}, b),
 \be\label{kul}
\check R_{b}(u)=P_5+\frac{q^2y^2-1}{q^2-y^2
}P_{3}-\frac{1-q^3y^2}{q^3-y^2}P_1,\qquad y=q^{-u}.
 \ee

For some special values of $q$, namely, if $q^3=1$,  there is
another solution too
\bea g(u)=\frac{-1+q^4 q^{up}}{q^3-q q^{up}},\qquad
q^3=1.\label{qroot1}\eea

These all solutions satisfy the remaining equations also, which
don't give additional constraints.

The solution (\ref{sal}) is given in H.~Saleur's paper
 \cite{s}. This formula can be checked also via the
general solution, defined by (\ref{rij}). In the classical limit
$q \rightarrow 1$ it can be represented in the common form
\be \check{R}_a(u)=a(u)I+b(u)P,\ee
 with $I,\; P$
unit and permutation operators. The other solution (\ref{kul}),
which we called as Kulish's solution, as it corresponds to the $R$
operator discussed in the articles \cite{k2,kp_m}, in the
classical limit $q \rightarrow 1$ have the form \cite{k1}
\be \check{R}_b(u)=a(u)I+b(u)P+c(u)K, \ee
 where $K$ is an operator, which together
with unit and permutation operators, commute with $osp(1|2)$
superalgebra generators.

Two solutions have the same braid group limit
$R^{\pm}=\mathrm{lim}_{u\to \pm\infty}R(u)$. $R^{+}$ corresponds
to the formula (\ref{runiv-q}), derived for the fundamental
representation. $R^{-}=P (R^{+})^{-1}P$.

At the point $q^{2u_0}=-q^{-1}$ the solution (\ref{sal}) reduces
to the higher spin projector, ${\check R}_{a}(u_0)\approx P_5$.
Therefore via a "fusion" procedure, developed in \cite{krs}, it is
possible to generate the solution of the YBE for the higher
dimensional representations of $osp_q(1|2)$ superalgebra, using
solution ${\check R}_{a}(u)$ (\ref{sal}). The ${\check R}_{a}(u)$
matrix itself could be obtained by the fusion from the $4\times 4$
matrices $R(u)$, defined by (\ref{rmat})
($\check{R}(u_0=\lambda)\approx P_3$).

For the solution (\ref{kul}) there is no such point $u_0$, for
which $\check R_{b}(u_0)$ would become proportional to the maximal
$P_5$ projection operator, so it is impossible to apply the fusion
method for finding the solutions for YBE with $R$ defined on the
higher spin representations. We shall discuss the consequences of
it in more detail in the subsection \ref{uks}. Note, that at the
poles $q^{u'}=q^{-1}$ and $q^{u''}=q^{-3/2}$ of (\ref{kul}) the
performing of the "fusion" procedure would reproduce the
fundamental $\check{R}_{b}$ or would decrease the dimension of the
quantum space to $1$.

Let us recall now the exceptional solution (\ref{qroot1}). Such
solutions to YBE for the cases when $q$ is a root of unity can
meet also for the higher dimensional representations (see the
subsection \ref{uks}). As it is known from the analysis of the
quantum (super-)algebras \cite{s1,kkh}, the structure of the set
of the non reducible representations and their fusion rules are
deformed, when $q$ is a root of unity, so the cases, like to the
(\ref{qroot1}) require separate investigation.

\subsection{Uniqueness of the solutions \label{uks}}
 We have seen that there are two general solutions to the
 Yang-Baxter equation (\ref{YBR1}),
when $1,2,3$ refer to the $3$-dimensional fundamental irreps. Only
one of them (\ref{sal}) allows to construct descendant solutions
 for higher spin
representations and belongs to the class, defined by equation
(\ref{rij}). But is it enough to conclude that there are no other
solutions for higher spin representations, besides them?

We can try to check it for some simple cases. At first let us
write the most general form of $\check{R}^{(44)}(u)$, satisfying
 (\ref{YBR1}), if $1,2,3$ are $4$ dimensional irreps:
\be \check{R}^{(44)}(u)=P_7+f(u) P_5+g(u) P_3+ h(u)P_1. \ee
Inserting it in the YBE, via not so hard analysis, we can separate
equations, where there are only $f(u),\; f(v),\; f(u+v)$
functions. Actually there is one independent equation of this
kind, namely
\bea f(v)=\frac{(f(u)-1)q^3+f(u+v)\left(1-(1+f(u)q^3+q^6)\right)}
{(f(u+v)-1)q^3+f(u)\left(1-(1+f(u+v)q^3+q^6)\right)}. \eea
And we can find the solution for $f(u)$, which is unique (more
precisely there is always freedom of the rescaling of the spectral
parameter $u\to pu$, with arbitrary $p$)
\bea f(u)=\frac{1+q^{3+u}}{q^3+q^{u}}. \eea
Then discussing the equations, where only  $f(u),\; f(v),\;
f(u+v)$ and $g(u),\; g(v),\; g(u+v)$ are involved, we find two
possible solutions for the $g(u)$
\bea g(u)=
\frac{(1+q^{3+u})(q^{2+u}-1)}{(q^3+q^{u})(q^2-q^u)}\quad
\mbox{(a)},\qquad g(u)=\frac{q^{5+u}-1}{q^5-q^{u}}\quad
\mbox{(b)}.\label{gs}\eea

 After analyzing the equations,
which contain also $h(u),\; h(v),\; h(u+v)$, we find that only one
solution $g(u)$ (\ref{gs}, a) is consistent with YBE. And $h(u)$
function turns out to be
\bea
h(u)=\frac{(q^{3+u}+1)(q^{2+u}-1)(q^{u+1}+1)}{(q^3+q^{u})(q^2-q^u)(q+q^u)}.
\eea
And certainly, this is the solution belonging to the class of
mentioned solutions, i.e. it is possible by fusion (descendant)
procedure to obtain this solution from the product of the
$R^{(22)}$ matrices (\ref{rmat}), or by (\ref{rij}) relations.

There can be some special solutions also, for the exceptional
values of $q$, such as (\ref{qroot1}), but we shall not
concentrate our attention on them now.

We can go further and try to find solution for $5$-dimensional
representations also, looking for the solution in the form
\be \check{R}^{(55)}(u)=P_9+f(u) P_7+g(u) P_5+ h(u)P_3+e(u)P_1.
\ee
And quite similarly to the previous analysis, we can separately
discuss four kind of the equations, which contain correspondingly
only the functions $f$, the functions $f,\;g$, the functions
$f,\;g,\;h$ and at last  all functions $f,\;g\;,h\;e$. The first
type of equations contain only one independent equation:
\bea
f(v)=\frac{f(u+v)\left(q^8+(1+f(u)q^4+1\right)+\left(1-f(u)\right)q^4}
{f(u)\left(q^8+(1+f(u+v)q^4+1\right)+\left(1-f(u+v)\right)q^4}.
\eea
 It has solution
\bea f(u)=\frac{q^4-q^u}{q^{4+u}-1}. \eea
The only arbitrariness here is regarding to the rescaling of the
spectral parameter $u$ - $u\to(p u)$, which leaves the YBE
invariant.

 Then the
second type of the equations gives two possible solutions for
$g(u)$.
\bea g(u)=\frac{(q^4-q^u)(q^3+q^u)}{(q^{4+u}-1)(1+q^{3+u})} \quad
\mbox{(a)}, \qquad g(u)=\frac{q^7-q^u}{q^{7+u}-1}\quad
\mbox{(b)}.\label{g1} \eea

 And the
third kind of the equations gives the unique solution for $h(u)$
with general $q$,
\bea
h(u)=\frac{(q^4-q^u)(q^3+q^u)(q^2-q^u)}{(q^{4+u}-1)(1+q^{3+u})(q^{2+u}-1)},
\eea
corresponding to the one of the $g(u)$ solutions, namely to
(\ref{g1}, a).

 For the
second one, (\ref{g1}, b), there is no $h(u)$ function, which
could satisfy to all the equations. So from the three kind of the
equations we find out the $f(u),\;g(u),\;h(u)$ functions, and the
fourth kind of the equations gives us the unique solution of the
function $e(u)$,
 \bea
e(u)=\frac{(q^4-q^u)(q^3+q^u)(q^2-q^u)(q+q^u)}{(q^{4+u}-1)
(1+q^{3+u})(q^{2+u}-1)(q^{1+u}+1)},
 \eea
which together with the other functions satisfy to all the YB
equations. This solution is also consistent with the formula
(\ref{rij}).

Note, that in the case of the fundamental $3$-dimensional irreps,
there are only two kind of the equations in the YBE set, i.e,
equations which contain only the $f(u)$ function, and the
equations with both of the $f(u)$ and $g(u)$ functions. So, after
finding two possible solutions for the $g(u)$ function, there are
no additional equations, which could exclude the second solution
of $g(u)$. Hence both of the two solutions (\ref{sal}, \ref{kul})
are valid.

And at the end we would like to clarify, whether there is a
$R^{(3r)}$ generalization for the (\ref{kul}) $R_b$-matrix. We can
try to find solution (if there is any) for the YBE (\ref{YBRk}),
where the $1,2$ spaces are the fundamental three dimensional
irreps, and the space $3$ is the five-dimensional irrep, when the
scattering of the three-dimensional irreps is described by the
solution (\ref{kul}).
\bea
\check{R}^{(33)}_{b}(u-v)\check{R}^{(35)}(u)\check{R}^{(35)}(v)
=\check{R}^{(35)}(v)\check{R}^{(35)}(u)\check{R}^{(33)}_{b}(u-v).
\eea
 We write the
$\check{R}^{(35)}$ in this form
\bea \check{R}^{(35)}=\check{P}_7+f(u)\check{P}_5+g(u)\breve{P}_3,
\eea
where the projectors $\check{P}_r$ are acting on the $V_3\otimes
V_5\to V_5\otimes V_3$.
 And we can verify by straightforward calculations
 that there is no solution $\check{R}^{(35)}$ for the YBE of this
 kind. So, we can conclude, that there is no general
 universal $R^{(rp)}(u)$-matrix, which would satisfy to the
 YBE, being $osp_q(1|2)$-invariant, with $R^{(33)}(u)$ defined in (\ref{kul})
 as it's particular case.

\section{ $ RLL=LLR $ relations with $9\times 9$ $R$-matrices;
the rational limit. } \setcounter{equation}{0}

Relying on the fusion procedure \cite{krs}, as it is stated
already, it is possible to obtain the solution (\ref{sal}) to the
YBE for the three dimensional representations of $osp_q(1|2)$
superalgebra, starting from the solution (\ref{rmat}) for the
two-dimensional ones. Similarly we can try to find the Lax
operator, with the three dimensional auxiliary space from the
fusion of two Lax operators with two-dimensional auxiliary spaces,
defined in the previous sections.

The operator $P_3 L^{(2 r)}(u)L^{(2r)}(u-u_0)$ with the point
$u_0=\lambda$, where $\check{R}(u_0)\approx P_3$ (\ref{rmat}),
serves as a such operator. It can be proved by repeatedly applying
the YB equations. Then by some reformulations we can find the ($3
\times 3$) matrix representation of the Lax operator. Let us
demonstrate it step by step.

From the YBE (\ref{rll}) it follows that
\be(-1)^{p_{c_1}(p_{b_2}+p_{c_2})}{(P_3)}_{b_1 b_2}^{a_1 a_2}
L_{c_1}^{b_1}(u)L_{c_2}^{b_2}(u-u_0)= L_{b_2}^{a_2}(u-u_0)
L_{b_1}^{a_1}(u){(P_3)}_{c_1 c_2}^{b_1
b_2}(-1)^{p_{b_2}(p_{a_1}+p_{b_1})}.\ee
So the operator-matrix
\be(PLL)_{c_1 c_2}^{a_1 a_2
}(u)=(-1)^{p_{c_1}(p_{b_2}+p_{c_2})}{(P_3)}_{b_1 b_2}^{a_1 a_2}
L_{c_1}^{b_1}(u)L_{c_2}^{b_2}(u-u_0), \label{sah1}\ee
 which acts as matrix
on $V_2\otimes V_2=V_1+V_3$ auxiliary space, really is nonzero
only in the three dimensional space $V_3$. The orthonormalized
eigenvectors of the projector $P_3$ are $\{1,0,0,0\}$,
$\{0,\frac{i\sqrt{q}}{\sqrt{1-q}},\frac{1}{\sqrt{1-q}},0\}$,
$\{0,0,0,1\}$. By means of the $3\times 4$ matrix constructed by
them
\be\mathbb{V}=\left(\ba{cccc}1&0&0&0\\
0&\frac{i\sqrt{q}}{\sqrt{1-q}}&
\frac{1}{\sqrt{1-q}}&0\\
0&0&0&1\ea\right),\ee
 and by the following transformation
\be L^{(3)}(u)=\mathbb{V}\cdot (P L L)(u)\cdot \mathbb{V}^{\tau},
\label{sah2}\ee
we arrive at an three dimensional matrix-operator $L^{(3)}(u)$,
which satisfies to the YBE
$$R^{(33)}(u-v)L^{(3)}(u)L^{(3)}(v)=L^{(3)}(v)L^{(3)}(u)R^{(33)}(u-v)$$
defined on the spaces $V_3\otimes V_3\otimes V_{arbitrary}$, with
$9\times 9$ matrix $R^{(33)}(u)$. It is obtained by fusion from
the tensor product of the $4\times 4$ matrices $R(u)$
(\ref{rmat}): on the space $V_2^{a}\otimes V_2^{a'}\otimes
V_2^{b}\otimes V_2^{b'}$ it is
\be R^{(33)}(u)=(\mathbb{V}P_{3}^{a a'}\;\mathbb{V}P_{3}^{b
b'})\cdot R_{ab'}(u+ u_0)R_{a'b'}(u)R_{ab}(u)R_{a'b}(u-u_0)\cdot
(\mathbb{V}^{\tau}P_{3}^{a a'}\; \mathbb{V}^{\tau}P_{3}^{bb'}).\ee
 The last matrix coincides with the solution (\ref{sal}) given in the previous
Section. There can be some uncertainties, caused by the matrix
representations of the projector operators. {\textit{It is
obtained by the redefinition of the matrix representations of the
algebra generators, taking place during the fusion procedure}.}
Note, that in general we can define a matrix
$U=\left(\ba{ccc}1&0&0\\
0&x_1&0\\
0&0&-x_1 x_2\ea\right)$, which transforms the three dimensional
representations of the algebra generators
$a$ to $U^{-1}\cdot a \cdot U $, particularly $f=\left(\ba{ccc}0&0&0\\
x_1&0&0\\
0& x_2&0\ea\right)$ generator transforms into $\left(\ba{ccc}0&0&0\\
1&0&0\\
0&-1&0\ea\right)$. It is equivalent to the automorphism $f\to
k^{-x} f k^{y},\;\;e\to k^{-y} e k^{x}$, with $q^x=x_2,\;q^y=x_1$.
So the more general form of the $3 \times 3$ Lax operator is
defined by $U^{-1}\cdot L^{(3)}(u) \cdot U$, which is
\bea
\mathbf{L}(u)=\left(\ba{ccc}\mathbf{L}_1^1&\mathbf{L}_1^2&\mathbf{L}_1^3\\
\mathbf{L}_2^1&\mathbf{L}_2^2&\mathbf{L}_2^3\\
\mathbf{L}_3^1& \mathbf{L}_3^2&\mathbf{L}_3^3\ea\right)(u),
\label{sah3}\eea
with matrix elements (as the finding of the Lax operator with
conventional fundamental auxiliary space is reasonable, when the
quantum spaces are also conventional odd dimensional
representations, with with eigenvalues of $h$ being integers,
below we take everywhere $(-1)^{2 h}=1$)
\bea\nn &&\mathbf{L}_1^1(u)=(-1)^h[-1/2 + a_x + h + 2 u
]_q-\frac{\alpha\sqrt{q}}{1+q},\\\nn
&&\mathbf{L}_1^2(u)=\frac{\sqrt{1-q}g_{10}x_1}{q^2-1}\left((i)^h
q^{\frac{h}{2}+\frac{1}{2}+a_x+2 u}+(i)^{-h}
q^{\frac{1}{2}-\frac{h}{2}}\alpha\right)f,\\\nn&&
\mathbf{L}_1^3(u)=\frac{ q^{1/2 + a_x +2 u}x_1 x_2
g_{10}^2}{q^2-1}f^2,
\\\nn&&\mathbf{L}_2^1(u)=\frac{-i\sqrt{1-q}(1+q)c_{10}}{g_{10}x_1}
\left((i)^h q^{\frac{h}{2}-1}\alpha-(i)^{-h} q^{-
a_x-\frac{h}{2}-2u}\right)e,
\\&&\mathbf{L}_2^2(u)=\left((-1)^h[-1/2 + a_x + 2 u
]_q-\frac{1+q}{\sqrt{q}}\alpha f e+\alpha[h-1/2]_q\right)c_{10},
\\\nn&&\mathbf{L}_2^3(u)=\frac{i\sqrt{1-q}
c_{10}g_{10}x_2}{q^2-1}\left((i)^{-h} q^{1+\frac{h}{2}}\alpha-
(i)^h q^{-\frac{h}{2}+a_x+2 u}\right)f,
\\\nn&& \mathbf{L}_3^1(u)=\frac{(1-q)(1+q)^3 c_{10}^2q^{-3/2 - a_x -2
u}}{x_1 x_2 g_{10}^2}e^2,
\\\nn&&\mathbf{L}_3^2(u)=\frac{\sqrt{1-q}(1+q) c_{10}^2}{g_{10}x_2}
\left((i)^h q^{-\frac{h}{2}-\frac{1}{2}}\alpha+(i)^{-h}q^{-
a_x-\frac{1}{2}+\frac{h}{2}-2u}\right)e,
\\\nn&&\mathbf{L}_3^3(u)=\left((-1)^{-h}[-1/2 + a_x - h + 2 u
]_q-\frac{\alpha\sqrt{q}}{1+q}\right)c_{10}^2, \eea
where $a_x$ is a number, defined by $a_{20}=\alpha q^{-a_x}$,
$\alpha=\pm 1$. The $a_{20},\; c_{10},\;g_{10}$ arbitrary
constants come from the definition of $2\times 2$ operator
(\ref{lgg}). Although, as it was mentioned before, by using the
super-algebra's automorphism, shift of the spectral parameter and
redefinition of the co-products, it is possible to fix all the
constants, we preferred to represent more general form of the Lax
operator. Remind, that there is also an arbitrariness of
multiplication by a function, which in above formulas is fixed, by
choosing appropriate $a_1(u)$ function.

As we see this matrix-operator has a well defined classical limit,
$q\to 1$, with additional requirements
\be x_1=x_{10}\sqrt{1-q},\;\;\; x_2=x_{20}\sqrt{1-q}.\ee
 If define
$q=e^{\it{\mathbf{h}}}$, then at the limit ${\it{\mathbf{h}}}\to
0$, we arrive at
\bea\nn &&\mathbf{L}_1^1(u)=(-1)^h( h-\frac{1}{2}+ a_x + 2 u
)-\frac{\alpha}{2},\qquad\mathbf{L}_1^2(u)=\frac{-g_{10}x_{10}}{2}
\left((i)^h +(i)^{-h} \alpha\right)f,\\\nn
&&\mathbf{L}_1^3(u)=\frac{-x_{10} x_{20} g_{10}^2}{
2}f^2,\qquad\qquad\qquad\quad\quad\;\;\;
\mathbf{L}_2^1(u)=\frac{-2 i c_{10}}{g_{10}x_{10}} \left((i)^h
\alpha-(i)^{-h} \right)e,\\
&&\qquad\qquad\mathbf{L}_2^2(u)=\left((-1)^h( a_x-\frac{1}{2} + 2 u )-2\alpha f e+\alpha(h-1/2)\right)c_{10},\\
\nn&&\mathbf{L}_2^3(u)=\frac{-i c_{10}g_{10}x_{20}}{2}
\left((i)^{-h} \alpha- (i)^h \right)f,\qquad \mathbf{L}_3^1(u)=\frac{8 c_{10}^2}{x_{10} x_{20} g_{10}^2}e^2,\\
\nn&&\mathbf{L}_3^2(u)=\frac{2 c_{10}^2}{g_{10}x_{20}} \left((i)^h
\alpha+(i)^{-h}\right)e,\qquad
\mathbf{L}_3^3(u)=\left((-1)^{-h}(a_x - h -\frac{1}{2}+ 2 u
)-\frac{\alpha}{2}\right)c_{10}^2.\eea
Let us fix $a_x=\frac{1}{2}$ and $\alpha=1$, $c_{10}=1,\;\;
g_{10}=2,\;x_{10}=x_{20}=1$.
 \bea \mathbf{L}(u)=\!\left(\ba{ccc}(-1)^h(
h+ 2 u )-\frac{1}{2} &- 2\cos{[\frac{\pi}{2}h]}f&-2 f^2\\
2 \sin{[\frac{\pi}{2} h]} e&(-1)^h( 2 u )-2 f e+(h-\frac{1}{2}
)&-2\sin{[\frac{\pi}{2} h]}
f\\
2 e^2&  2\cos{[\frac{\pi}{2}h]} e&(-1)^{-h}( 2 u-h
)-\frac{1}{2}\ea\right).\eea
In the case of choice $x_{10}=\sqrt{-i},\;x_{20}=-\sqrt{i}$, which
corresponds to a real matrix representations of the $9\times 9$
$R(u)$ matrix, we have
\bea \mathbf{L}(u)=\!\left(\ba{ccc}(-1)^h(
h+ 2 u )-\frac{1}{2} &- 2\sqrt{-i}\cos{[\frac{\pi}{2}h]}f& 2 f^2\\
2\sqrt{i} \sin{[\frac{\pi}{2} h]} e&(-1)^h( 2 u )-2 f
e+(h-\frac{1}{2} )&2\sqrt{i}\sin{[\frac{\pi}{2} h]}
f\\
-2 e^2& 2\sqrt{-i}\cos{[\frac{\pi}{2}h]} e&(-1)^{-h}(  2 u-h
)-\frac{1}{2}\ea\right). \eea

\section{Discussion}

The investigation of the quantum extension of the super-algebra
$osp(1|4)$ (may be, it is true for the general case $osp(1|2n)$,
too), shows, that in the quantum deformation case, besides of the
conventional irreducible representations, there are irreps, which
don't appear in the classical case. Suppose the irrep with minimal
dimension in the quantum case don't coincide with the conventional
fundamental representation. We can find the solutions to YBE for
that minimal representation, and then, from the fusion procedure,
find all the solutions of YBE with arbitrary representations. As
we have checked, the minimal representation of the $osp_q(1|4)$
has four dimension, while the fundamental representation is a
five-dimensional one.

We expect that analogous relations hold in the general case
$osp_q(1|2n)$ too. There are works (see \cite{z} and references
therein), which show connection between the finite dimensional
representations of the $osp_q(1|2n)$ and  the non-spinorial
representations of the quantum deformation of $so(2n+1)$ algebra.
It is expected that there are also non conventional "spinorial"
representations for the $osp_q(1|2n)$ algebra, with complex
eigenvalues of the $h_i$ Cartan matrices, as it was in the case of
the $osp_q(1|2)$ algebra (the even dimensional representations).
And surely then the minimal dimensional irreducible
representations belong to the "spinorial" kind. Although such
representations have no classical limit (when $q \to 1$), it is
sensible to construct Lax operator defined on the auxiliary space
with such minimal dimensional representation. The Lax operator
with auxiliary space being the conventional fundamental
representation can be constructed by a fusion procedure from the
simpler operators, like in the case $n=1$ discussed in this paper.

\section*{\large{\bf{Acknowledgements}}}

D.~K. and Sh.~Kh. thank the Volkswagen Foundation for partial
financial support. D.K. is grateful to Leipzig University for
hospitality and to Deutsche Forschungsgemeinschaft for supporting
his visit.

\end{document}